# Intrinsic chemical and structural inhomogeneity in lightly doped $La_{1-x}Sr_xMnO_3$

Tomohiro Shibata,[a] Bruce Bunker,[a] John Mitchell,[b] and Peter Schiffer [c]

a) Department of Physics, University of Notre Dame, Notre Dame, IN 46556

b) Materials Science Division, Argonne National Laboratory, Argonne, IL 60439

c) Department of Physics and Materials Research Institute, Pennsylvania State University, University Park PA 16802

**Abstract**

X-ray absorption fine structure measurements of the Sr and La K edges of the solid solution $La_{1-x}Sr_xMnO_3$ reveal a consistent deviation from a random distribution of Sr at the La/Sr sites for $x \lesssim 0.3$. Local structural disorder on the cation sublattice in the low-$x$ samples is also observed to differ in the vicinity of the La-rich and Sr-rich clusters. The local clustering and structural disorder establish an intrinsic chemical as well as structural inhomogeneity on the nanometer scale, which may provide a mechanism for the nucleation of magnetoelectronic phase separation.

PACS numbers: 75.30.-m, 75.30.Vn 61.10.Ht, 87.64.Fb

There has been much recent interest in the relation among the structural, magnetic, and transport properties of perovskite manganites with the general formula $Ln_{1-x}R_xMnO_3$, where "Ln" is a lanthanide and "R" is an alkaline earth. The lanthanide and alkaline earth ions form a solid solution on the so-called A-site of the perovskite, thus doping these materials to yield a mixed-valent $Mn^{3+}/Mn^{4+}$ lattice. Because of this mixed-valent state, these materials display a number of remarkable properties including an anomalously large negative magnetoresistance, the so-called "colossal" magnetoresistance (CMR) effect [1]. Importantly, the compelling physics of these materials is driven by a close coupling among lattice, electronic, and magnetic degrees of freedom. This coupling has recently been shown through a wide range of experimental techniques and theoretical treatments to result in electronic phase separation between different magnetoelectronic states at low temperatures [2]. These studies strongly suggest that the phase separation is an intrinsic property of the manganites and not attributable to poor sample quality or large-scale compositional inhomogeneity.

Despite the importance of large-scale electronic inhomogeneity to the physics of the manganites, there has not yet been a detailed study of how the rare-earth and alkaline earth A-site cations distribute themselves in the local atomic structure of these materials. Although these compounds are intrinsically disordered due to the combination of these cations, most theoretical studies have assumed that the material is microscopically homogeneous; that is, the ions are distributed randomly. In this Letter we report a detailed XAFS study of the local structure of one of the canonical manganite materials, $La_{1-x}Sr_xMnO_3$. We find that there is an *intrinsic* nanoscale clustering of the Sr ions in this material for $x \lesssim 0.3$. We further show that such clustering impacts the local electronic structure through modification of the cation sublattice. Such chemical and structural inhomogeneities may in turn influence the nucleation of the observed large-scale magnetoelectronic phase segregation in the manganite compounds [3].

One of the reasons we chose to study the canonical manganite $La_{1-x}Sr_xMnO_3$ is because the photoelectron backscattering amplitudes and phase shifts of La and Sr are quite different—allowing them to be easily differentiated in XAFS analysis. Samples of $La_{1-x}Sr_xMnO_{3+\delta}$ were prepared by two different methods. The first method ($x$=0.025, 0.075, 0.225, 0.325, 0.425, 0.475) involved a solution-based preparation of a coprecipitated precursor powder, which was then fired at 1000 $^o$C and then rapidly cooled (< 1 min) to room temperature in the processing atmosphere [4]. The second method (x=0.05, 0.10, 0.175, 0.3, 0.375, 0.45) was a standard solid state route starting from $MnO_2$, $SrCO_3$, and $La_2O_3$ and involving repeated high temperature firing and grinding. In each case the final firing was typically at 1400 $^o$C for 24 hours followed by furnace cooling over several hours. The lightly doped perovskite manganites are known to be oxygen hyperstoichiometric unless appropriate processing parameters are observed [5]. In both of the methods used here, a processing atmosphere was chosen to produce oxygen stoichiometric material. The oxygen content of samples with $x \leq 0.175$ was verified by triplicate iodometric titrations, and we found $\delta = 0.00(1)$ for each.

The x-ray absorption fine-structure spectroscopy (XAFS) technique reveals local structural information about different atomic species in a sample. This technique has proven useful in the study of non-random site occupation in alloys – allowing the determination of interatomic distances, variation in these distances, i.e. Debye Waller factors (DWF), and types and numbers of neighboring atoms within the first few coordination shells of the x-ray excited atom [6,7]. By measuring spectra about the x-ray absorption edges of the various constituent atoms, this information may be separately obtained about the La, Sr, or Mn ions in $La_{1-x}Sr_xMnO_3$.

The XAFS spectra were measured using fluorescence detection at the Sr and La K edges (16104 eV and 38925 eV, respectively) at the MRCAT undulator beamline at the Advanced Photon Source [8]. A harmonic-reduction mirror was used, and both incident and fluorescence intensities were monitored by gas-filled ion chambers. The samples were attached to a cold finger of a DISPLEX refrigerator and were measured at 10K. Three scans were repeated for each sample to verify repeatability. The extraction of the XAFS oscillations $\chi(k)$ (as a function of photoelectron wave number $k$) followed standard procedures [6]. Although there is in principle a concentration-dependent amplitude reduction of fluorescence spectra, comparison with transmission data shows the correction to be small for these samples.

The Fourier transform of $k^2$-weighted data $k^2\chi(k)$ ( $\tilde{\chi}(r)$ ) for Sr K edge data is shown in Fig. 1 for the various samples. The overall spectra are qualitatively similar for the Sr and La edges as expected from the known structure. For further analysis, the FEFFIT program [9] was used to fit $\tilde{\chi}(r)$ to a model of the local environment constructed by the calculation of the XAFS spectra using FEFF6 [10]. The fitting allows us to determine the interatomic distances, the numbers of neighbors in the surrounding shells, and DWF within the context of the known structure [11,12].

One of the most important results of the fitting is the Sr occupancy around the Sr site ($x_{Sr}^{Sr}$). If we treat $x_{Sr}^{Sr}$ as a free parameter in the fits, we find that the data for small $x$ can be much better fit by values of $x_{Sr}^{Sr}$ greater than would be expected if the Sr ions were randomly distributed on the cation sublattice. This is demonstrated in Fig 2, where we plot the real and the imaginary part of $\tilde{\chi}(r)$ for x = 0.10 at the Sr K edge (gray solid line with circle). We show fits to these data from two different fit methods: the dotted line is with $x_{Sr}^{Sr}$=0.24 (the best fit value) while the dashed line is with $x_{Sr}^{Sr}$=0.10 (expected from a random distribution of Sr). The peak of the amplitude (solid line) at R ≈ 3.65Å [13] corresponds to a La/Sr shell with slight spectral interference from a Mn shell peaked at R ≈ 2.9Å [13]. Due to this interference, a significant difference between these two fits appears around R ≈ 3.1 - 3.5Å in the phase of $\tilde{\chi}(r)$ and the fit with $x_{Sr}^{Sr}$=0.24 gives much better than $x_{Sr}^{Sr}$ =0.10, which is manifested by the discrepancy from the data,

$$\Delta\tilde{\chi}(r) = \sqrt{[\Delta\text{Re}(\tilde{\chi})]^2 + [\Delta\text{Im}(\tilde{\chi})]^2}$$, plotted for both of these models in the lower panel of the figure.

The best fit values of $x_{Sr}^{Sr}$ as a function of $x$ are shown in Fig. 3. If the Sr and La atoms were distributed randomly on the A-site of the perovskite, the curve should fall on the dashed line in Fig. 3. This is the case for x > 0.3, but a consistent deviation from a random occupancy is observed for $x \leq 0.3$, indicating a tendency for Sr clustering. Note that these results are not explained by simple alloy fluctuations: If the alloy were random, there would naturally be some regions that show a higher Sr-Sr (and La-La) correlation, but these would be averagedwith other regions that have *anti-correlations* in the near-neighbor occupancy. Although most of our samples were powders, for $x = 0.30$ we compared a powder and crushed single-crystal sample grown in a floating-zone furnace and obtained similar values of $x_{Sr}^{Sr} = 0.38\pm0.04$ and $x_{Sr}^{Sr} = 0.34\pm0.04$ respectively. This result and the insensitivity to the two vastly different methods used for synthesizing powders argue strongly for an intrinsic origin to the Sr clustering that is not dependent on the details of sample preparation. While it is known from local structure studies [14,15] that structural inhomogeneity of the Jahn-Teller distorted $MnO_6$ octahedra persists up to x~0.35, the cation clustering observed here provides a heretofore unidentified degree of chemical and structural inhomogeneity.

The observed difference in La-Mn and Sr-Mn distances on the cation sublattice is consistent with an observed inhomogeneity in the Mn oxidation state distribution, offering corroborative evidence of the clustering of Sr and La sites *via* its structural impact. A striking concentration dependence of the spectra shown in Fig. 1 appears for the Sr-Mn coordination at R ≈ 2.9Å [13]. For x ≳ 0.175, a sharp Gaussian-shaped peak with nearly constant intensity is observed, but the intensity of this peak becomes smaller and broader, and actually develops structure for $x \leq 0.175$. Although not shown here, the La-edge spectra display the same behavior in La-Mn bond lengths. This behavior indicates a large increase in local structural disorder in the cation sublattice at lower Sr concentrations. This increase in structural disorder becomes even more apparent in the fitting results for the Sr-Mn and La-Mn distances, shown in Fig. 4. As shown in this figure, the Sr-Mn and La-Mn distances are relatively constant above $x \approx 0.175$, but split into a few different distances below that point. The $x < 0.175$ results are as expected, given the structural phase transition at that concentration, and our results agree with the known orthorhombically distorted crystal structure of the endpoint compound, LaMnO$_3$. Note however that the Sr-Mn distances exhibit a consistently smaller splitting than the La-Mn distances, indicating that the structural disorder is smaller in the vicinity of the clustered Sr ions. The single observed distance at higher concentrations indicates that the cation sublattices of these compounds show less structural disorder. From the La/Sr-Mn data alone we cannot, however, comment on the persistence of Jahn-Teller distortions into the metallic regime that have been reported from neutron pair distribution function analysis [14].

Cation disorder in manganites has previously been approached phenomenologically in two ways: (1) assuming an average structure gauged through an average A-site radius, $<r_A>$ [16], or (2) by introducing the variance of A-site cation radius to account for large size mismatch not captured by the average size alone [17]. The former approach has shown success in qualitatively understanding how cation size impacts the Curie temperature, $T_C$. The latter approach has shown a linear dependence of $T_C$ on the cation variance which can be explained phenomenologically through local deviations of structure (Jahn-Teller distortion, Mn-O-Mn bond angles, etc.) that in turn impact the global electronic structure. Our results offer corroborative evidence for this latter phenomenology, since the clustering of Sr ions demonstrates that the average radius cannot adequately explain the microscopic electronic structure. Furthermore, the varying Sr-Mn and La-Mn environments demonstrate convincingly that the global electronic structure is decidedly inhomogeneous.

The existence of local structural inhomogeneity has strong implications for the physics of the manganites, since almost all theoretical treatments of these materials have assumed homogeneity on the length scale of a few atomic bonds. The inhomogeneity we observe is certainly relevant to magnetoelectronic phase separation, which has been shown to exist on length scales from a few nanometers to several microns. While we expect the clustering of Sr ions to be on a nanometer length scale, since Coulomb energetics would prevent larger clusters from forming, such small clusters could act as nucleation sites for the larger-scale phase separation [3]. Thus the presence (or absence) of such clustering may help explain differences in phase separation between different manganite compositions. These results may also have repercussions beyond the manganites: For example, recent measurements of the magnetic properties of ruthenates may also be explained by Sr clustering. [18]

In conclusion, we observe a tendency for Sr clustering in lightly doped $La_{1-x}Sr_xMnO_3$. The local structural disorder in the vicinity of the clustered Sr ions is also significantly smaller than in the vicinity of clustered La ions, presumably attributable to a higher average $Mn^{3+}$ concentration in the latter regions. While we have only studied a single cation pair (La and Sr), one must presume that similar local inhomogeneity exists in other compositions. Such inhomogeneity should therefore be considered in any local theoretical treatment of these compounds and particularly in the context of magnetoelectronic phase separation.

This work has been sponsored by the US Department of Energy (DoE), Office of Science under Contract No. W-31-109-ENG-38 and by NSF under grant numbers DMR-9701548 and DMR-0101318. The MRCAT is supported by the DoE under Contract DE-FG02-94-ER45525 and the member institutions. Use of the APS was supported by the DoE, under Contract No. W-31-102-Eng-38.

**Figure Captions**

**Figure 1.** Fourier transforms ($\tilde{\chi}(r)$) of the data $k^2\chi(k)$ for alloy samples at the Sr K edge. The data range $k$ = 3-14 Å$^{-1}$ was transformed. The ordinate has been shifted for easier display.

**Figure 2.** The data and fit results for $\tilde{\chi}(r)$ at the Sr K edge shown for the x = 0.10 sample (k = 3 - 15 Å$^{-1}$ are transformed). The real and imaginary parts of the spectrum (gray solid line with circles) are shown along with the best fit for $x_{Sr}^{Sr}$ ($x_{Sr}^{Sr}$ = 0.24, dotted line) and with a fit fixing $x_{Sr}^{Sr}$ to the value expected from a random distribution ($x_{Sr}^{Sr}$ =0.10, dashed line). The peak of the amplitude (solid line) at R ≈ 3.65Å [13] corresponds to a La/Sr shell with a slight spectral interference with a Mn shell peaked at R≈ 2.9Å. The lower panel shows the difference between the data and the fits, $\Delta\tilde{\chi}(r) = \sqrt{[\Delta\text{Re}(\tilde{\chi})]^2 + [\Delta\text{Im}(\tilde{\chi})]^2}$. Clearly, the larger $x_{Sr}^{Sr}$ results in a much higher quality fit to the data.

**Figure 3.** Third-shell Sr coordination about Sr atoms from the data fitting. If the site occupancy for these atoms were random, the results would be a linear function of $x$, as shown by the dashed line. The excess Sr at low doping indicates a tendency towards Sr-Sr (and thus also La-La) clustering.

**Figure 4.** The Sr-Mn and La-Mn distances as a function of $x$ at 10 K. For x ≥ 0.175, a single distance was used for both Sr-Mn and La-Mn bonds for the fit. Due to the increased distortion of the lattice for smaller x, three different distances for La-Mn (approximated as two short, two long and four intermediate distances) and two for Sr-Mn (six short and two

long distances) were used. This is due to a smaller local distortion of the lattice around Sr than La, not resolved into three different lengths for Sr-Mn due to a smaller splitting of the peaks (not due to experimental resolution.) The diffraction data are from references 5, 19, and 20.  The phase boundary between orthorhombic and rhombohedral (hatched) is from reference 21. Note that references 20 and 21 are room temperature measurements.

12. It is known that there are some systematic differences between FEFF calculations and experimental backscattering amplitudes and phase shifts (F. Bridges, Personal Communication). For this work, a sample of cubic perovskite $SrMnO_3$ was also measured and a correction factor determined. The change in our fitting results due to the correction factor was significantly smaller than the quoted uncertainty.

13. Note that peak positions in the Fourier transform in XAFS generally appear slightly shorter than actual bond lengths [6]. The real bond lengths for Sr-Mn and Sr-La/Sr bonds are $R \approx 3.4$ Å and $R \approx 3.9$ Å respectively.

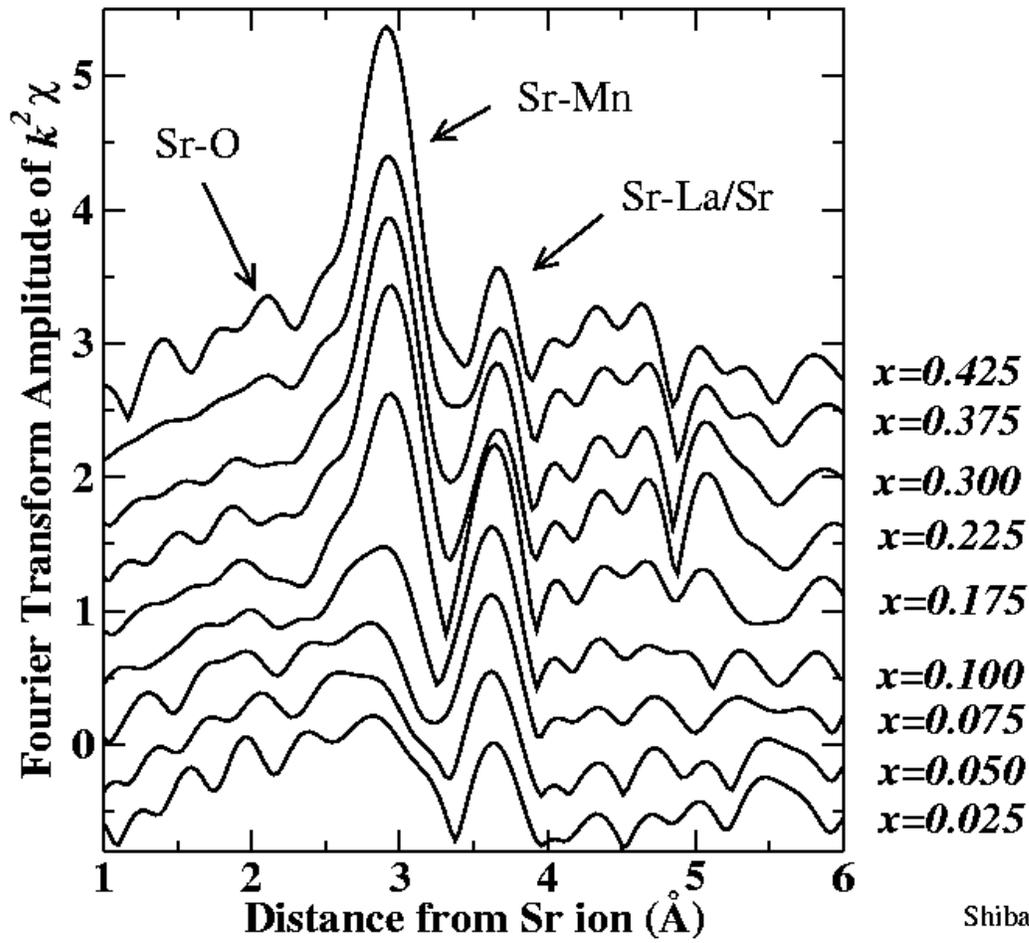

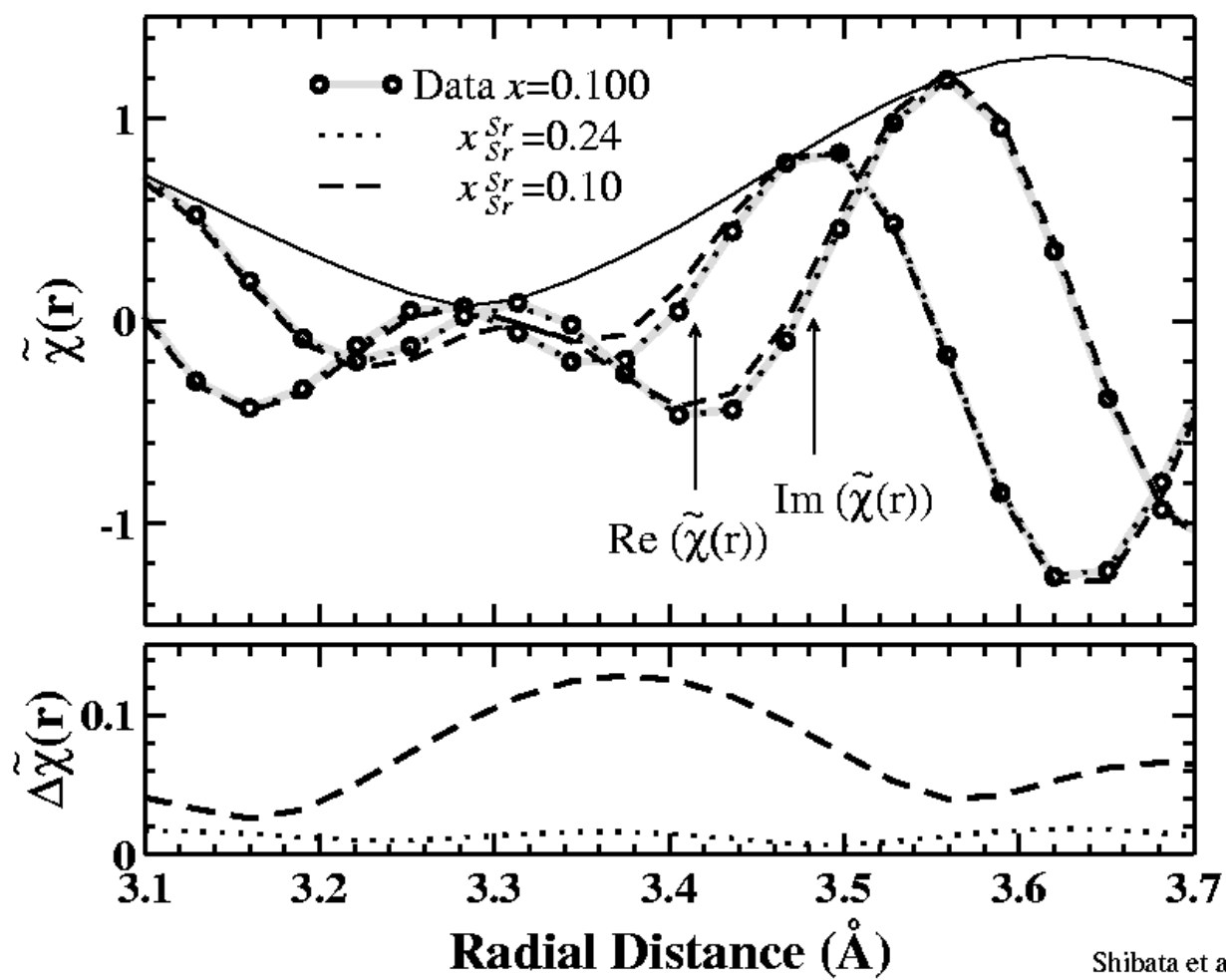

Shibata et al. Fig. 2

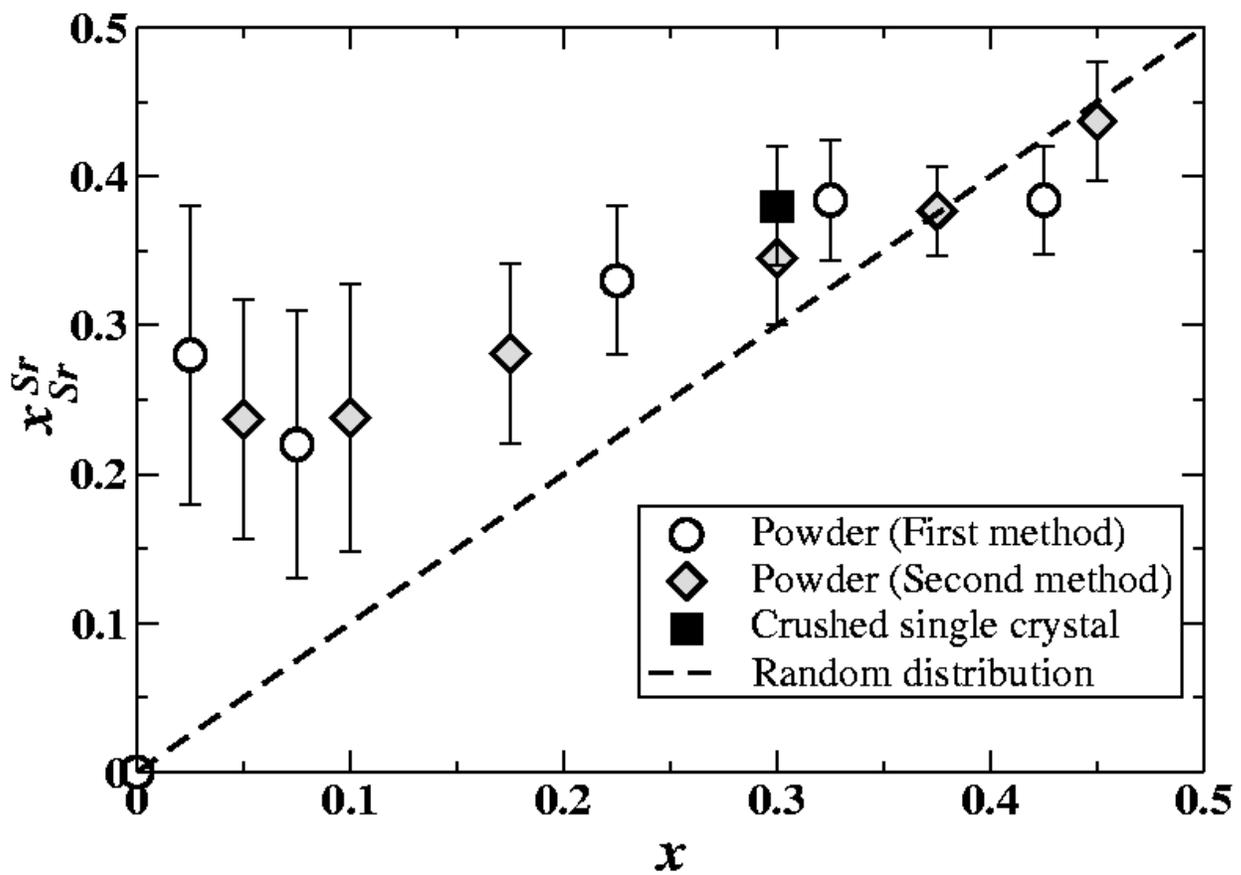

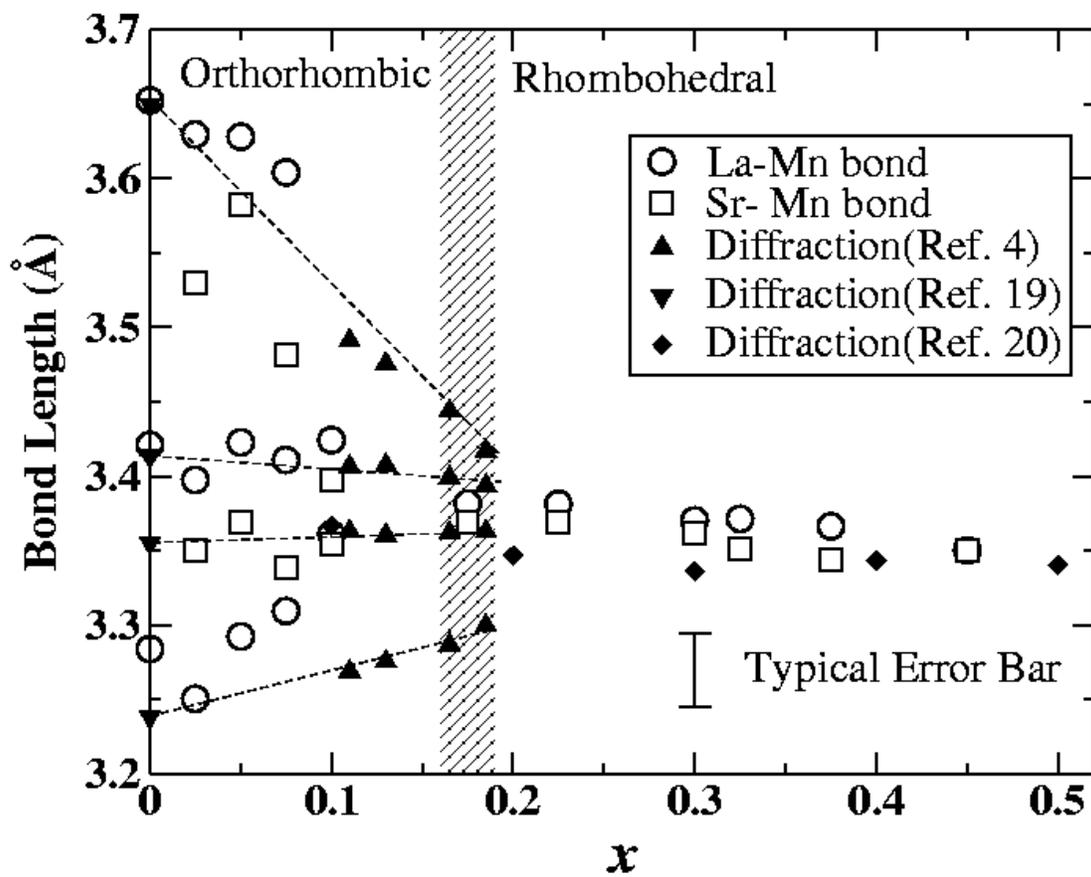